\documentclass[letter,twocolumn]{jpsj2} 

\def\runauthor{Hirono {\sc Kaneyasu-Fukazawa} and Kosaku {\sc Yamada}}

\title
{Theory on Superconductivity of
 CeIn$_3$ in Heavy Fermion System
}
	
\author
{ 
Hirono \textsc{(Kaneyasu-)Fukazawa}$^{1}$\thanks{E-mail address:
hirono@sci.u-hyogo.ac.jp} and Kosaku \textsc{Yamada}$^{2}$
}

\inst
{$^{1}$ Faculty of Science, Himeji Institute of
Technology (Department of Material Science, University of Hyogo),\\ 
Kamigori, Akou, Hyogo 678-1297 \\
$^{2}$ Department of Physics, Kyoto University, Kyoto 606-8244
}

\recdate
{
\today
}

\abst
{
On the basis of a three-dimensional Hubbard model,
the superconducting mechanism of CeIn$_3$ under high pressure is investigated
by the third-order perturbation theory with respect to the on-site
Coulomb interaction $U$. Here we propose the $d$-wave pairing state induced
by antiferromagnetic spin fluctuations. The estimated superconducting
transition temperature is lower by one order than that in the
two-dimensional system for the same value of $U/W$ ($W$=bandwidth).
This result is consistent with the difference between CeIn$_3$ and CeRhIn$_5$.
}

\kword
{
superconductivity, CeIn$_3$, CeRhIn$_5$, dimensionality, heavy fermion system, perturbation theory
}

\begin{document}
\sloppy
\maketitle

Heavy fermion compound CeIn$_3$ \cite{CeIn3} has been confirmed to exhibit
the properties of unconventional superconductivity. For instance, in the
${}^{115}$In-NQR measurement \cite{CeIn3-NQR}, no coherence peak has
been observed in the temperature dependence of $1/T_{1}$. The
superconducting (SC) state appears under the pressure $P$ with a
critical value $P_{\rm c}=2.55$ GPa, and the maximum value of the
SC-transition temperature $T_{\rm c}^{\rm max}=0.2$ K. The SC state exists
near the antiferromagnetic (AF) phase with the ordering vector
$Q=(\pi,\pi,\pi)$ at Ce atoms \cite{CeIn3-AF}. 
The situation strongly suggests 
that superconductivity should be connected to 
three-dimensional (3D)-AF spin fluctuations.

In this paper, we assume that the CeIn$_3$ system is given by the
Fermi liquid state at low temperatures and study the SC mechanism
induced by the wave number dependence in the effective interaction
between quasi particles originating from the Coulomb interaction among
$f$-electrons. The SC state is considered to be realized on the main Fermi surface of 
Ce $4f$-electrons\cite{CeIn3-Band}. 
We pursue a possibility of the $d$-wave pairing state due to AF spin
fluctuations near ${\bf Q}=(\pi,\pi,\pi)$. To describe the heavy
fermions, here a 3D Hubbard model is adopted and the effective
interaction is evaluated on the basis of the third-order perturbation
theory (TOPT) in terms of the on-site Coulomb interaction
\cite{Hotta}\cite{Jujo}. It is concluded that the SC mechanism of CeIn$_3$ is $d$-wave pairing induced by 3D-AF spin fluctuations near ${\bf Q}$=$(\pi,\pi,\pi)$.

In previous papers\cite{Arita,Takimoto},
two-dimensional (2D) and 3D single-band Hubbard models were generally
investigated within fluctuation exchange approximation (FLEX). Although
they treated different AF spin fluctuations depending on
dimensions, they obtained the same result that $T_c$ in the 3D system
is lower than that in the 2D system.
In addition to the characteristics of the AF spin fluctuation, we treated
different wave number dependences due to the vertex correction in 2D and 3D dimensions\cite{JPSJ}. 

In this paper, we focus on CeIn$_3$ to discuss the microscopic
mechanism of superconductivity. On the basis of TOPT, we investigate the
effect of the interaction with the complex wave number dependence on
the SC mechanism in 3D CeIn$_3$.
The calculation based on TOPT includes also the normal self energy in
this study of CeIn$_3$. We treat the wave number dependence
originating from the main Fermi surface of CeIn$_3$ and the 
 calculated $T_{\rm c}$ is reduced by renormalized Fermi energy $E_{F}$. As a result, we performed a new analysis of CeIn$_3$ in detail on the basis of TOPT.
Next, we clarify the suppression of $T_c$ by calculating the
dependence of  $T_{\rm c}^{\rm TOPT}/T_{\rm c}^{\rm RPA-like}$ on
dimensionality. The change in $T_c$ between 2D and 3D systems applies to
that between 2D CeRhIn$_5$ and 3D CeIn$_3$. With regard to CeTIn$_5$ (T=Co, Rh
and Ir), Takimoto, Hotta and Ueda\cite{THM} studied the SC mechanism
with a 2D orbital-degenerate Hubbard model. They indicated
the appearance of the $d$-wave superconductivity next to the AF magnetic
phase in the case where the crystal splitting energy is large. On the
other hand, Nisikawa, Ikeda and Yamada\cite{NIY} investigated the
$d$-wave superconductivity in CeIr$_x$Co$_{1-x}$In$_5$ with the 2D single-band Hubbard model. 

Here, we apply the single-band model to both CeIn$_3$ and CeRhIn$_5$
due to
the following reason. The main Fermi surface of CeIn$_3$ is a single
band as shown by the band calculation of Betsuyaku\cite{CeIn3-Band}. The
3D single band induces the wave number dependence of the AF spin
fluctuation, which plays the main role in superconductivity. Both
CeIn$_3$ and CeRhIn$_5$ have the same kind of Ce compounds. Therefore, we
apply TOPT to the superconductivity in the single band of
CeRhIn$_5$ as well as CeIn$_3$. The main difference between the two materials is
the dimensionality of the Fermi surface. We clarify the effect
of dimensionality on the wave number dependence in the analysis based on TOPT, and moreover, we show the justification of the quasi-2D model in CeRhIn$_5$.

We explain the formulation in the following.
The Hubbard Hamiltonian is given by
\begin{equation}
{\cal H}=
-t_1 \sum_{{\bf i},{\bf a},\sigma} c^{\dag}_{{\bf i}\sigma}c_{{\bf i+a}\sigma}
+t_2 \sum_{{\bf i},{\bf b},\sigma} c^{\dag}_{{\bf i}\sigma}c_{{\bf i+b}\sigma}
+U \sum_{{\bf i}}n_{{\bf i}\uparrow} n_{{\bf i}\downarrow},
\end{equation}
where $c_{{\bf i}\sigma}$ is an annihilation operator for a quasi
particle with spin $\sigma$ at site ${\bf i}$, ${\bf a}$ and ${\bf i}$
are, respectively, the vectors connecting nearest-neighbor and
next-nearest-neighbor sites in a simple cubic lattice. Transfer
integrals $t_1$ and $t_2$ denote nearest-neighbor and
next-nearest-neighbor hopping amplitudes, respectively. $U$ is the
on-site Coulomb interaction, and $n_{{\bf
i}\sigma}$=$c^{\dagger}_{{\bf i}\sigma}c_{{\bf i}\sigma}$. Here, we
consider that the parameters $t_1$, $t_2$ and $U$ include a common renormalization factor\cite{NIY} for constructing the heavy fermion. As shown in Fig. 3, we adjust the dispersion $E_{\bf k}$ so as to reproduce the main Fermi surface of heavy fermions constructed mainly by Ce $4f$-electrons in the simple cubic structure\cite{CeIn3-Band},
leading to
\begin{equation*}
\hspace{-3.2cm}E_{\bf k}=-2t_1(\cos k_x+\cos k_y+\cos k_z)
\end{equation*}
\begin{equation}
+4t_2(\cos k_x \cos k_y+ \cos k_y \cos k_z+\cos k_z \cos k_x). 
\end{equation}
Then, we obtain the bare Green's function of the quasi particle
as $G_0(k)$=$1/[i\omega_n-(E_k-\mu_0)]$, where $k$ is a short-hand notation defined as $k$=$({\bf k},\omega_n)$, ${\bf k}$ is momentum and $\omega_n=\pi T(2n+1)$ is the fermion Matsubara frequency with temperature $T$. 
Note that the chemical potential $\mu_0$ for the non-interaction case is determined by the electron number $n$ (per site and spin) as $n$=$\sum_{k}G_0(k)$, where $\sum_k$=$(T/N)\sum_{\bf k}\sum_n$ and $N$ is the number of sites. 
The dressed normal Green's function $G(k)$ is given by $G(k)$=$1/[i\omega_n-(E_k-\mu)-\Sigma_{\rm n}(k)]$, where $\Sigma_{\rm n}(k)$ is the normal self-energy, given by TOPT with respect to $U$ as 

\begin{figure}
\begin{center}
\includegraphics[width=0.3\textwidth]{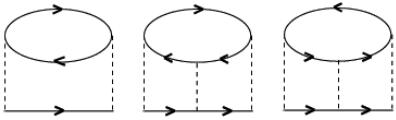}
\end{center}
\caption{Diagram of the normal self-energy $\Sigma_{\rm n}(q, \omega_{\rm n})$. The solid line is the bare Green's function
$G_0$. The broken line is the Coulomb interaction $U$. The broken line 
of $U$ connects only solid lines possessing opposite spins.}
\label{fig:1}
\end{figure}
\begin{figure}
\begin{center}
\includegraphics[width=0.45\textwidth]{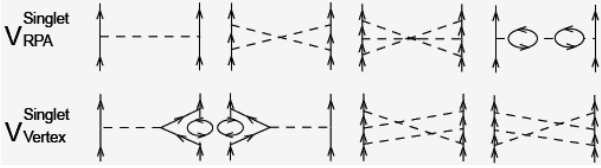}
\end{center}
\caption{Diagrams for the effective interaction of singlet pairing within the third-order perturbation with respect to $U$. The solid line is the bare Green's function $G_0$. The broken line is the Coulomb interaction $U$. The broken line of $U$ connects only solid lines possessing opposite spins. The two external lines have opposite spins. The effective interaction is divided into the RPA-like part and the vertex correction. The latter begins with the third-order terms.}
\label{fig:2}
\end{figure}
\begin{equation*}
\hspace{-7.3cm}\Sigma_{\rm n}(k)=
\end{equation*}
\begin{equation}
\sum_{k'}\{ U^2\chi_0(k-k')+U^3[\chi_0^2(k-k')+\phi_0^2(k+k')] \}G_0(k'),
\end{equation}
with 
\begin{equation}
\chi_0(q)=-\sum_{k}G_0(k)G_0(q+k),
\end{equation}
\begin{equation}
\phi_0(q)=-\sum_{k}G_0(k)G_0(q-k).
\end{equation} 
Here, $q$ denotes a short-hand notation $q$=$({\bf q},\nu_n)$, where $\nu_n$=$2\pi T n$ is the boson Matsubara frequency. Note that the chemical potential $\mu$, shifted from $\mu_0$, is again determined by the condition $n$=$\sum_{k}G(k)$. We show the diagram of the normal self energy in Fig. 1. 

An effective pairing interaction $V$ between quasi particles is evaluated using TOPT. Although the origin of superconductivity is investigated by total terms in $V$, in order to analyze the role of $V$ in detail, it is convenient to divide it into two parts as 
\begin{equation}
V(k,k')=V_{\rm RPA}(k,k')+V_{\rm vertex}(k,k'),
\end{equation}
where $V_{\rm RPA}$ represents the terms obtained by the random phase
approximation (RPA) and $V_{\rm vertex}$ indicates the third-order
vertex correction terms. The RPA-like term reflects the nature of simple spin fluctuations, while the third-order vertex correction terms originate from the electron correlation other than the spin fluctuations. For singlet pairing, $V_{\rm RPA}$ and $V_{\rm vertex}$ are given by
\begin{equation}
{\it V}_{\rm RPA}(k,k')=U+U^2\chi_0(k-k)+2U^3\chi_0^2(k-k'),
\end{equation}
\begin{equation*}
\hspace{-6.2cm}{\it V}_{\rm Vertex}(k,k')= 
\end{equation*}
\begin{equation}
2 U^3 Re \sum_{k''}G_0(k+k''-k')(\chi_0(k+k'')-\phi_0(k+k''))G_0(k'') .
\end{equation}
We show the diagrams for singlet pairing in Fig. 2. 

An anomalous self-energy $\Sigma_{\rm a}$ 
is expressed using $V(k,k')$ and an anomalous Green's function $F(k)$ as $\Sigma_{\rm a}(k)$=$-\Sigma_{k'}V(k,k')F(k')$.
 At $T$=$T_{\rm c}$, the linearized Eliashberg equation including $\Sigma_{\rm a}$ and $F(k)$ is reduced to the eigenvalue equation, 
\begin{equation}
\lambda\Sigma_{\rm a}^{\dagger}(k)=-\sum_{k'}V(k,k')|G(k')|^2\Sigma_{\rm a}^{\dagger}(k').
\end{equation}
When the eigenvalue $\lambda$ becomes unity, the SC state is realized and $T_{\rm c}$ is obtained. 
We solve the equation on the assumption that $\Sigma_{\rm a}^{\dagger}$ has singlet or triplet pairing symmetry. 
We divide the first Brillouin zone into 64$\times $64$\times $64 momentum meshes and take $N_{\rm f}$ = 512 for Matsubara frequency $\omega_n$. The bandwidth $W$ ($W^{3D}\sim 12 t_1$) is a necessary range of $\omega_n$ for reliable calculations. The range is covered with the condition: $|W|< \pi T N_{\rm f}$. To satisfy the condition, we calculate in the region with $T>$0.0037.

As a result, the dominant symmetries are $d_{x^2-y^2}$- and
$d_{3z^2-r^2}$-wave pairings, which are degenerate due to the space symmetry of the cubic system. In Fig. 4, we show the wave number dependence of the anomalous self-energy $\Sigma^A(q,\omega_n$ = $\pi T)$ for $d_{x^2-y^2}$-wave pairing.
On the other hand, we did not obtain stable solution in the present calculations for other pairing symmetries such as $d_{xy}$-wave.

We explain in detail the mechanism of the $d$-wave pairing, which indicates
$d_{x^2-y^2}$ or $d_{3z^2-r^2}$-wave pairing.
To describe the main large-volume Fermi surface \cite{CeIn3-Band} (see Fig. 3), we choose a parameter set as  $t_2=-0.2$ and $n$=0.45, near the half-filling $n$=0.5.
The main Fermi surface with a nesting property enhances the bare susceptibility
$\chi_0({\bf q},\nu_n)$, due to the feature of 3D-AF spin fluctuations
near ${\bf Q}$=$(\pi,\pi,\pi)$, as shown in Fig. 3.
The AF spin fluctuation originating from the RPA-like term provides an advantageous contribution to the eigenvalue $\lambda$ 
for the $d$-wave pairing, as shown in Fig. 5           . 
In the case including only the RPA-like terms, 
$\lambda$ always increase with increasing $U$. However, 
it is significantly suppressed for a large $U$ value
when all terms in TOPT are taken into account,
since the vertex correction terms suppress the $d$-wave superconductivity.

Next, we show the $U$ dependence of $T_{\rm c}$ in Fig. 6. 
Here, it is emphasized that the evaluated $T_{\rm c}$ in the unit 
of $t_1$ is consistent with the experimental SC-transition temperature
$T_c$=0.2 K in the value $T_{\rm c}\sim$ 0.003. We estimated the
renormalized $T_c$ with the electron effective mass $m^*$ obtained by
the experiment of dHvA. The details of the 
estimation are as the follows.
The renormalized bandwidth $W$ is given as $zW_0$ with the
renormalization factor $z$.  As another relation, $W$ is given 
as $W=12t_1$ from the $t_2=-0.2$. Here, $W_0$ is a bare bandwidth. The renormalization factor is defined by $z=m_0/m^*$ and $m_0$ is the bare electron mass. 
dispersion $E_{\bf k}$ of eq. (2) at
\begin{figure}
\begin{center}
\includegraphics[height=4.6cm]{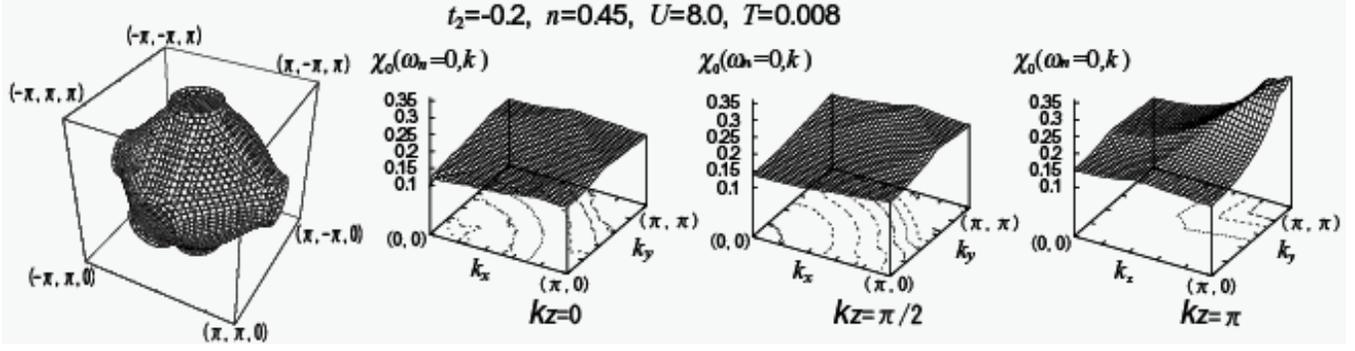}
\end{center}
\caption{Fermi surface in a quarter first Brillouin zone and bare susceptibility $\chi_0(q, \omega_n=0)$.}
\label{fig:3}
\end{figure}
\begin{figure}
\begin{center}
\vspace{5.4cm}
\includegraphics[height=10.4cm]{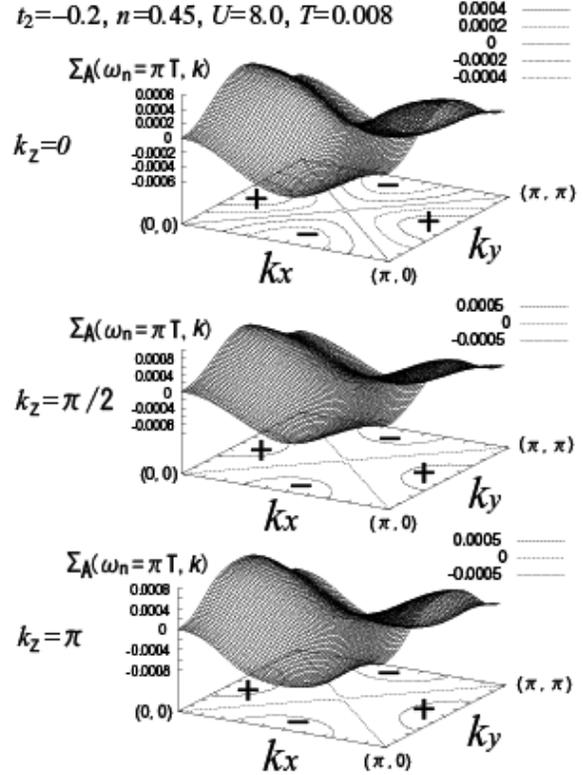}
\end{center}
\caption{Anomalous self-energy $\Sigma^A(q,\omega_n$ = $\pi T)$ for $d_{x^2-y^2}$-wave pairing.}
\label{fig:4}
\end{figure}
Therefore, the relation of $W$ is $12t_1=zW_0$. The bare bandwidth $W_0$ is $W_0 \sim 0.1$ Ry=$1.58 \times 10^4$ K in the band
calculation\cite{CeIn3-Band}. $z$ is obtained as $z=1/16$ using the
cyclotron mass of dHvA\cite{dHvA}. From the relation, the renormalized
$t_1$ is obtained as $t_1=0.819\times 10^2$ K. By means of $t_1$, the
calculated value $T_c\sim$0.003($t_1$) at $U\sim$9.0 corresponds to
$T_c$=0.246 K, which is near the experimental SC-transition temperature
$T_c$=0.2 K. Thus, the present calculation for $T_{\rm c}$ well explains
the possibility of the $d$-wave pairing state in CeIn$_3$. Because the
calculated $T_c$ on the basis of TOPT is near the experimental
SC-transition temperature, we consider that the system has a strong
electron correlation $U\sim9.0$, which is larger than the value of the bandwidth.

Furthermore, it is quite instructive to consider the comparison
between 2D and 3D cases. We investigate the effect of dimensionality on
the constant strength of the substantial electron correlation. Here, the
strength of the substantial electron correlation means the value of $U$
in comparison with the bandwidth. Thus, $U/W$ keeps a constant value in the change from 2D to 3D systems.
For the same value of
$U/W\sim$ 3/4
($t_2=-0.2$, $n=0.45$, $W^{\rm 3D} \sim 12 t_1$ and $W^{\rm 2D} \sim 8 t_1$),
we obtain $T_{\rm c}^{\rm 2D}$=0.042 in a 2D square lattice 
and $T_{\rm c}^{\rm 3D}$=0.0067 in a 3D simple cubic lattice.
Namely, $T_{\rm c}$ for the 3D system is lower by about one order
than that for the 2D system.
This is consistent with the experimental findings that
$T_{\rm c}^{\rm 3D}\approx$0.2 K for CeIn$_3$ (3D system)
and $T_{\rm c}^{\rm 2D}\approx$2.1 K for CeRhIn$_5$ (quasi-2D system)
\cite{CeRhIn5}. 

We show in Fig. 7 the change in
$T_{\rm c}^{\rm TOPT}/T_{\rm c}^{\rm RPA}$ between 2D and 3D systems
obtained by the present calculation using the dispersion
\begin{equation*}
\hspace{-2.4cm}E_{\bf k}=-2t_1(\cos k_x+\cos k_y+t_z\cos k_z)
\end{equation*}
\begin{equation}
+4t_2(\cos k_x \cos k_y+ t_z\cos k_y \cos k_z+t_z\cos k_z \cos k_x).
\end{equation}
Here, the hopping integrals $t_1$ and $t_2$ in the c-axis direction
are multiplied by $t_z$.
Namely, the dispersion relations with $t_z$=0 and 1 correspond
to those for the 2D square and 3D simple cubic lattices, respectively.
As a result, the suppression of $T_{\rm c}$ by the vertex correction is
stronger in the 3D system than in the 2D system. By including the suppression on
the basis of TOPT, the difference in $T_{\rm c}$ becomes one order
between 2D and 3D systems. 
In addition to this, the decrease in $T_c$ is
slight with the increase in the 3D characteristic in the region near the
quasi-2D system. Actually, in the range that $t_z$ is small ($t_z=0 \sim
0.4$), the change in $T_c$ by TOPT is small under the condition that $U/W$
is constant (see the inset of Fig. 7). Therefore, the $T_{\rm c}$ calculated
on the basis of the 2D model is appropriate for CeRhIn$_5$, even if the band
structure of CeRhIn$_5$ possesses a weak 3D characteristic.

\begin{figure}
\begin{center}
\includegraphics[height=5.4cm]{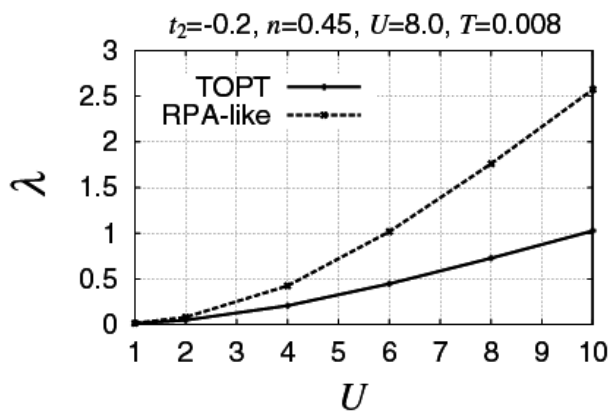}
\end{center}
\caption{$U$ dependence of the eigenvalue $\lambda$ obtained by TOPT and RPA-like terms.}
\label{fig:5}
\end{figure}

\begin{figure}
\begin{center}
\includegraphics[height=5.5cm]{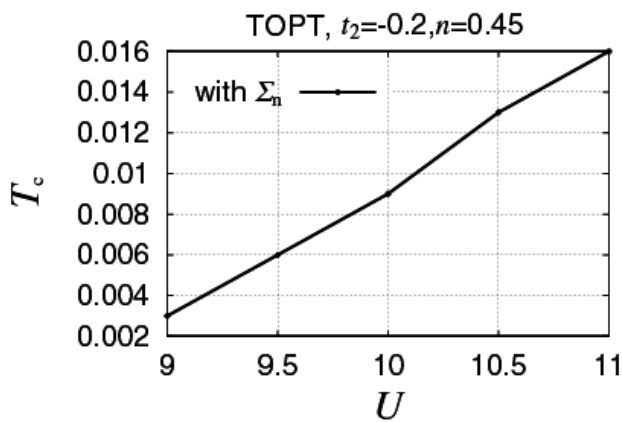}
\end{center}
\caption{$U$ dependence of $T_{\rm c}$ calculated by TOPT.}
\label{fig:6}
\end{figure}

Here, we explain the different suppression depending on dimensionality
in the analysis of TOPT. We consider the wave number dependence of
$\chi_0$ on the $k_x$-$k_y$ plane perpendicular to the $k_z$-axis, as shown in Fig. 3.
In the quasi-2D system with AF spin fluctuations, the same prominent
peaks exist near $Q$=($\pi$, $\pi$) on planes at all $k_z$. In the 3D
system, $\chi_0$ has prominent peaks around $Q$=($\pi$,$\pi$) on
planes only near $k_z$=$\pi$. On the other hand, since the prominent
peaks exist on planes at all $k_z$ in the quasi-2D system, the
RPA-like term naturally gives high $T_c$ in the quasi-2D system rather
than in the 3D system. This fact has been known in previous studies \cite{Arita,Takimoto,Monthoux}.
On the other hand, the complex wave number dependence exists except
for the region around $Q$. As opposed to the prominent peaks, the
momentum space possessing the complex wave number dependence occupies
a larger region in the 3D system than in the quasi-2D system. For
example, in the 3D system, the complex wave number dependence spreads
on planes at $k_z=0$ and $\pi/2$ in Fig. 3. Therefore, the suppression
by the vertex correction in the 3D system is stronger than that in the quasi-2D system. 
In addition to the characteristics of the AF spin fluctuation, we obtain
$T_c$ by including the effect of the complex wave number dependence in
TOPT. The calculated $T_c$ is near to the realistic SC-transition
temperature by including each wave number dependence for each 
dimensional system. Therefore, we compare the $T_c$ obtained by TOPT with the experimental SC-transition temperature in CeIn$_3$ and CeRhIn$_5$.

\begin{figure}
\begin{center}
\includegraphics[height=6.95cm]{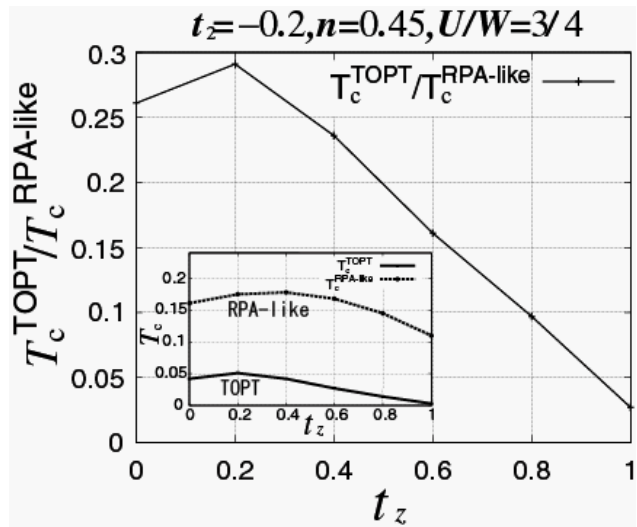}
\end{center}
\caption{Change in $T_{\rm c}^{\rm TOPT}/T_{\rm c}^{\rm RPA}$ between 2D and 3D systems at the same value of $W/U$. The inset shows $T_{\rm c}^{\rm TOPT}$ and $T_{\rm c}^{\rm RPA-like}$ obtained by TOPT and RPA-like calculations.}
\label{fig:7}
\end{figure}

We summarize our study of the superconductivity in CeIn$_3$.
By means of TOPT based on the 3D Hubbard model, we explained the mechanism of the superconductivity of CeIn$_3$ from a microscopic point of view.
It is concluded that the SC mechanism of CeIn$_3$ is $d_{x^2-y^2}$
(or $d_{3z^2-r^2}$-wave) pairing, mainly induced by 3D-AF spin fluctuations
near ${\bf Q}$=$(\pi,\pi,\pi)$. 
In the present calculation including the suppression of $T_{\rm c}$ by
the third-order vertex correction,
$T_{\rm c}$ in the 3D cubic systems is lower by one order than that
in the 2D square system for the same value of $U/W$,
in good agreement with the experimental results for 2D and 3D Ce-based 
heavy fermion superconductors, CeRhIn$_5$ and CeIn$_3$.
It has been pointed out by means of FLEX that the superconductivity induced by AF spin fluctuations
is suppressed in the 3D system compared with that in the 2D system
\cite{Arita}\cite{Takimoto},
but here we stress that the difference is obtained also by means of TOPT including the vertex correction in this paper.
Moreover, we found that the suppression of $T_{\rm c}$ by the vertex
correction is stronger in the 3D system than in the 2D system for the same value of $U/W$. 
A part of this paper is contained in the proceeding of ASR2002. \cite{ASR}
The authors thank Dr. H. Ikeda, Professo T. Hotta and Professer H. Harima for valuable discussions.


\begin{thebibliography}{99}
\bibitem{CeIn3} 
N. D. Mathur, F. M. Grosche, S. R. Julian, I. R. Walker, D. M. Freye, R. K. W. Haselwimmer and G. G. Lonzarich: Nature {\bf 394} (1998) 39.
\bibitem{CeIn3-NQR}
S. Kawasaki, T. Mito, G.-q. Zheng, C. Thessieu, Y. Kawasaki, K. Ishida, Y. Kitaoka, T. Muramatsu, T. C. Kobayashi, D. Aoki, S. Araki, Y. Haga, R. Settai and Y. Onuki: Phys. Rev. B {\bf 66} (2002) 054521.
\bibitem{CeIn3-AF} 
P. Morin, C. Vettier, J. Flouquet, M. Konczykowski, Y. Lassailly, J.-M. Mignot and U. Welp: Low. Temp. Physics {\bf 70} (1988) 377.
\bibitem{CeIn3-Band} 
K. Betsuyaku: Dr. Thesis, Osaka University (1999). 
\bibitem{Hotta}
T. Hotta: J. Phys. Soc. Jpn. {\bf 63} (1994) 4126.
\bibitem{Jujo}
T. Jujo, S. Koikegami and K. Yamada: J. Phys. Soc. Jpn. {\bf 68} (1999) 1331.
\bibitem{Arita}
R. Arita, K. Kuroki and H. Aoki: Phys. Rev. B {\bf 60} (1999) 14585.
\bibitem{Takimoto}
T. Takimoto and T. Moriya: Phys. Rev. B {\bf 66} (2002) 134516.
\bibitem{JPSJ}
H. Fukazawa, H. Ikeda and K. Yamada: J. Phys. Soc. Jpn. {\bf 68} (1999) 1331.
\bibitem{THM}
T. Takimoto, T. Hotta and K. Ueda: J. Phys. Condens. Matter. {\bf 14} (2002) L369.;
cond-mat/0212467.
\bibitem{NIY}
Y. Nisikawa, H. Ikeda and K. Yamada: J. Phys. Soc. Jpn. {\bf 71} (2002) 1140.
\bibitem{dHvA}
R. Settai, T. Ebihara, M. Takashita, H. Sugawara, N. Kimura,
K. Motoki, Y. Onuki, S. Uji and H. Aoki: J. Magn. Magn. Mater. {\bf 140-144} (1995) 1153.
\bibitem{CeRhIn5}
H. Hegger, C. Petrovic, E. G. Moshopoulou, M. F. Hundley, J. L. Sarrao, Z. Fisk and J. D. Thompson: Phys. Rev. Lett. {\bf 84} (2000) 4986.  
\bibitem{Monthoux}
P. Monthoux and G. G. Lonzarich: Phys. Rev. B {\bf 72} (2003) 884.
\bibitem{ASR}
H. Fukazawa and K. Yamada: J. Phys.: Condens. Matter {\bf 15} (2003) S2259.
\end{thebibliography}
\end{document}